\documentstyle[twocolumn,aps,epsfig]{revtex}
\begin{document}
\draft
\title{Super-Radiance and the Unstable Photon Oscillator}
\author{S. Sivasubramanian and A. Widom}
\address{Physics Department, Northeastern University, Boston MA 02115}
\author{Y.N. Srivastava}
\address{Physics Department, Northeastern University, Boston MA 02115} 
\address{Physics Department \& INFN, University of Perugia, Perugia Italy}
\maketitle

\begin{abstract}
If the damping of a simple harmonic oscillator from a thermally random 
force is sufficiently strong, then the oscillator may become unstable.
For a photon oscillator (radiatively damped by electric dipole moments), 
the instability leads to a low temperature Hepp-Lieb-Preparata super-radiant 
phase transition. The stable oscillator regime is described by the free 
energy of the conventional Casimir effect. The unstable  
(strongly damped) oscillator has a free energy corresponding to 
Dicke super-radiance. 
\end{abstract}

\narrowtext

\section {Introduction}

In many theoretical treatments of condensed matter systems, the 
Hamiltonian consists of a sum of kinetic energy and Coulomb potential  
energy. In reality, the transverse electromagnetic (radiation) fields 
are also important. If the transverse electromagnetic field is included 
in the description of condensed matter, then phenomena such as the Casimir 
effect\cite{1}, the Lamb shift\cite{2,3,4,5} and the long ranged 
inter-molecular interaction forces\cite{6,7,8} can be explained employing 
standard low order quantum electrodynamic perturbation theory.
If the coupling of the transverse electromagnetic field with matter 
is strong enough to produce an instability in the photon oscillators, 
then a transition occurs from a normal radiative phase to a  
super-radiant phase. The Dicke model\cite{9} has been studied as a 
prime example of this phase transition. In the Dicke model,
many two level molecules interact with a single photon mode of the 
radiation field. The Dicke model has been extensively studied by many 
workers\cite{10,11,12,13}, each adopting a somewhat different method to 
explain the physical properties of the system. Hepp and Lieb\cite{10}, 
analyzed the thermodynamic properties of the Dicke maser model and 
elucidated the mathematical structure of the super-radiant phase 
transition. 

In subsequent works of Wang and Hoe\cite{11} and Hepp and Lieb\cite{12},  
the coherent states of Glauber \cite{14} were employed for multi-photon 
oscillators in the Dicke model and the resulting phase transition has been 
investigated in great detail in these papers. 
The existence of such phase transitions in condensed matter 
systems has been investigated in a series of papers by Preparata and 
coworkers\cite {15,16,17,18,19,20}. Preparata has proposed 
several examples of super-radiant phases in a recent review\cite{18}. 
Water\cite{18,19,20} is a particularly interesting example. 

The purpose of this work is to present a very direct method for 
computing the thermodynamic region wherein super-radiant phase 
exists. When a condensed matter system interacts with the 
electromagnetic field, the field degrees of freedom may be 
described by a renormalized thermodynamic potential. For small 
oscillator displacements (in the normal 
radiation phase) the effective potential is parabolic. As the
thermodynamic parameters change, the effective oscillator frequency 
may turn imaginary. The resulting oscillator instability is the 
signature of the phase transition into a super-radiant state.

In Sec.\ II, the statistical thermodynamic method for calculating 
the effective potential of a photon oscillator mode will be discussed. 
The conditions for the stability of the oscillator will be exhibited. 
In Sec.\ III, the dynamics in real time of the oscillator will be 
discussed in terms of the photon mode propagator. The conditions for 
dynamic real time stability are shown to be the same as for 
thermodynamic stability. In Sec.\ IV, the example of 
a mesoscopic object placed in an optical cavity is considered. 
If the mesoscopic object has a sufficiently high polarizability, 
then a cavity mode can become unstable. In the Sec.\ V, the nature 
of the non-linear forces which stabilize the oscillator mode in 
the super-radiant phase are explored. The shift in the equilibrium 
position required to stabilize the oscillator in the low temperature 
super-radiant phase is exhibited. In the concluding Sec.\ VI, the 
notion of super-radiance with electric dipole coupling is related 
to the notion of coherent ferro-electricity. A material grain with a 
high electric polarizability, but otherwise without a net dipole 
moment, can be {\em induced} into a ferro-electric phase by an 
interaction with transverse electric fields. The transverse fields 
may induce a coherence in the electronic states then yielding a net 
electric dipole moment.
 
\section {Statistical Thermodynamics}

Let us consider a simple harmonic oscillator with  
momentum  \begin{math}{P} \end{math}, displacement 
\begin{math}{Q} \end{math} and 
frequency \begin{math}{\omega}_{\infty} \end{math}. 
The oscillator is coupled to a thermal bath described by 
the Hamiltonian \begin{math}{H} \end{math}. The free energy \cite{21} of 
the thermal bath \begin{math} {\cal F}_{0}  \end{math} is then 
determined by 
\begin{equation}
e^{-{\cal F}_0/k_B T}=Tr_{bath}\ e^{-H/k_B T}.
\end{equation}
The coupling between the thermal bath and the oscillator is described 
by a force  
\begin{math} f \end{math}. The total Hamiltonian is then 
\begin{equation}
{\cal H}={1 \over 2} \left( P^2 +\omega_{\infty}^{2}Q^2\right)+H-fQ.
\end{equation} 
which determines the total free energy 
\begin{math} {\cal F} \end{math} via 
\begin{equation}
e^{-{\cal F}/{k_B T}} = Tr\ e^{-{\cal H}/{k_B T}}
\end{equation}

If \begin{math} Q \end{math} and \begin{math} P \end{math} are classical  
to a sufficient degree of accuracy, then the the {\em total trace} may 
be approximated by 
\begin{equation}
Tr(...)=\int \int \left({dPdQ\over 2\pi \hbar }\right)Tr_{bath}(...).
\end{equation}
From Eqs.(1-4) one obtains 
$$
e^{-({\cal F}-{\cal F}_0)/{k_B T}}=
$$
\begin{equation}
\int \int \left({dPdQ\over 2\pi \hbar }\right)
e^{-P^2/2k_BT } e^{-{\cal G}(Q,T)/k_B T},
\end{equation}
i.e. 
\begin{equation}
e^{-({\cal F}-{\cal F}_0)/{k_B T}}=
\sqrt{{k_BT\over 2\pi \hbar^2}}\int dQ
e^{-{\cal G}(Q,T)/k_B T}, 
\end{equation}
where 
$$
{\cal G}(Q,T)=\left({\omega_\infty ^2 Q^2\over 2}\right)- 
$$
\begin{equation}
k_BT\ln
\left\{
{Tr_{bath}\ e^{-(H-fQ)/k_BT } \over Tr_{bath}\ e^{-H/k_BT}}
\right\}.
\end{equation}
Finally, if the force \begin{math} f \end{math} is {\em also classical} 
to a sufficient degree of accuracy, then 
\begin{equation}
\left\{
{Tr_{bath}\ e^{-(H-fQ)/k_BT } \over Tr_{bath}\ e^{-H/k_BT}}
\right\}\to \left<e^{fQ/k_BT}\right>,
\end{equation}
yielding 
\begin{equation}
{\cal G}_{classical}(Q,T)=(\omega_\infty ^2Q^2/2)-k_BT
\ln \left<e^{fQ/k_BT}\right>.
\end{equation}

The ``effective'' thermally induced potential of the oscillator is 
given by 
\begin{equation}
{\cal G}(Q,T)=\omega_\infty^2 (Q^2/2)
-k_BT\ln \left<e^{fQ/k_BT}\right> , 
\end{equation}
where (without loss of generality) one may assume 
(for \begin{math} Q=0  \end{math}) 
that \begin{math} \big<f\big>=0 \end{math}. For small values 
of the oscillator displacement, Eq.(10) implies 
\begin{equation}
\lim_{Q\to 0}({\cal G}(Q,T)/Q^2)= (\omega_0^2/2), 
\end{equation}
where, for a {\em classical force} \begin{math} f \end{math}, the 
shifted frequency \begin{math}\omega_0 \end{math} of the  
oscillator is given by
\begin{equation}
\omega_0^2= \omega_{\infty}^2 - \left<f^2\right>/k_B T . 
\end{equation}

The frequency shift (from \begin{math} \omega_\infty \end{math} 
downward to \begin{math} \omega_0  \end{math}) of the oscillator due 
to the interaction of the oscillator with the thermal bath is given 
by  
\begin{equation}
\omega_{\infty}^2-\omega_0^2 =\left<f^2\right>/k_B T .
\end{equation}
The oscillator is ``stable'' if the shifted frequency is real; i.e.
\begin{math} \omega_0^2>0  \end{math} or equivalently  
\begin{equation}
k_BT \omega_{\infty}^2 > \left<f^2\right>\ \ \ 
({\rm stable}).
\end{equation}
The oscillator is ``unstable'' if the shifted frequency is imaginary; 
i.e. \begin{math} \omega_0^2<0  \end{math} or equivalently  
\begin{equation}
k_BT \omega_{\infty}^2 < \left<f^2\right>\ \ \ 
({\rm unstable}).
\end{equation}

One may appreciate the dynamical significance of the shifted frequency 
by considering the spectral density 
\begin{math}S_{ff}(\omega) \end{math} of the force fluctuations 
\cite{22,23,24,25}
defined by 
\begin{equation}
\left<f(t)f(t^{\prime})\right> = \int_{-\infty}^{\infty} 
S_{ff}(\omega) \cos{\omega}(t-t^{\prime}) d{\omega}.
\end{equation}
For equal times  
\begin{equation}
\left<f^2\right>
= 2\int_0 ^{\infty} S_{ff}(\omega) d{\omega}
\end{equation}
The classical fluctuation response theorem asserts that 
\begin{equation}
S_{ff}(\omega) =\left({k_B T\over {\pi}}\right) 
\Re e\  \Gamma(\omega+i0^+)
\end{equation}
where \begin{math} \Gamma (\zeta )  \end{math} for 
\begin{math} \Im m \ \zeta >0 \end{math} is the dynamic damping 
coefficient for the oscillator. \begin{math} \Gamma (\zeta )  \end{math}  
will be discussed in detail in Sec.\ III below. Here we note that  
Eqs.(13), (17) and (18) imply that 
\begin{equation}
\omega_{\infty}^2 -\omega_0^2=\left({2\over {\pi}}\right)
\int_0 ^{\infty} \Re e\ \Gamma(\omega+i0^+)d{\omega}.
\end{equation}

If the integrated (over frequency) dissipative damping on the right 
hand side of Eq.(19) is sufficiently large, the the oscillator becomes 
unstable. The oscillator will be stable for  
\begin{equation}
\omega_\infty^2 > \left({2\over {\pi}}\right)\int_0 ^{\infty} 
\Re e\ \Gamma(\omega+i0^+)d{\omega}\ \ \ ({\rm stable}).
\end{equation} 
Eq.(20) has been here deduced by statistical thermodynamic reasoning. 
However, it is perhaps more simple to deduce Eq.(20) directly from 
the simple dynamical equations of motion for the oscillator.

\section {Real Time Oscillator Dynamics}

Let us now consider a simple harmonic oscillator with a canonical  
momentum  \begin{math}{P} \end{math}, displacement \begin{math}{Q} 
\end{math} and frequency \begin{math}{\omega}_{\infty } \end{math}. 
The oscillator is driven by a small time varying external force 
\begin{math} \delta f_{ext}(t) \end{math}.  
The total Hamiltonian, including the external force is then  
\begin{equation}
{\cal H}_{total}={\cal H}-\delta f_{ext}(t)Q
\end{equation}
where \begin{math} {\cal H} \end{math} is defined in Eq.(2).
In a phenomenological linear response theory, the small mean deviation 
from thermal equilibrium in the oscillator coordinate obeys the equation 
of motion
\begin{equation}
\delta \ddot{Q}(t)+\omega_\infty ^2\delta Q(t)=
\delta f(t)+\delta f_{ext}(t)
\end{equation}
where \begin{math} \delta f(t) \end{math} represents the induced change 
in the frictional damping force exerted on the coordinate by the thermal 
bath.

For simple engineering purposes, one often considers that the frictional 
damping force has a simple local (in time) form 
\begin{math} \delta f(t)=-\gamma \delta \dot{Q}(t) \end{math}. More 
generally, the damping force will be non-local in time; i.e.  
\begin{equation}
\delta f(t)=-\int^\infty _0 {\cal G}(s)\delta \dot{Q}(t-s)ds.
\end{equation} 

For complex frequency \begin{math} \zeta \end{math}, with  
\begin{math} \Im m \ \zeta > 0 \end{math}, if 
\begin{equation}
\delta f_{ext}(t)=\Re e \left\{ \delta f_{ext}(0)e^{-i \zeta t}\right\},
\end{equation}
then 
\begin{equation}
\delta Q(t)=\Re e \left\{D(\zeta)\delta f_{ext}(0)e^{-i\zeta t}\right\}
\end{equation} 
defines the dynamical oscillator propagator 
\begin{math} D(\zeta) \end{math}.
From Eqs.(23) and (25) 
\begin{equation}
\delta f(t)=\Re e \left\{\Pi (\zeta ) 
D(\zeta)\delta f_{ext}(0)e^{-i\zeta t}\right\},
\end{equation}
where 
\begin{equation}
\Pi (\zeta )=i\zeta \int_0^\infty {\cal G}(s)e^{i\zeta s}ds. 
=i\zeta \Gamma (\zeta ).
\end{equation}
From Eqs.(22), (25) and (26), one finds that 
\begin{equation}
D(\zeta )=
\left({1\over \omega_{\infty}^2 -\zeta^2 - \Pi (\zeta)}\right);
\end{equation}
i.e. \begin{math} \Pi(\zeta ) \end{math} is the ``self energy'' 
of the harmonic oscillator propagator. 

From the retarded (causal) nature \cite{26} of Eqs.(23) and (27) follows the 
dispersion relation
\begin{equation}
\Pi (\zeta)=\left({2 \over \pi}\right) \int_0^{\infty} {\omega \  \Im m\ 
\Pi (\omega+i0^+)\over(\omega^2 - \zeta^2)}d\omega
\end{equation}
Equivalently, in terms of the frequency dependent damping coefficient 
\begin{math} \Gamma (\omega +i0^+) \end{math},
\begin{equation}
\Im m\ \Pi (\omega+i0^+) =\omega \ \Re e\ \Gamma(\omega +i0^+)
\end{equation}
Eq.(29) reads 
\begin{equation}
\Pi (\zeta)=\left({2 \over \pi}\right) \int_0^{\infty} {\omega^2 \ 
\Re e\ \Gamma(\omega +i0^+)  )\over(\omega^2 - \zeta^2)}d\omega
\end{equation}

As \begin{math} \zeta \to 0 \end{math}, Eq.(25) reads as the thermal 
(static) response function 
\begin{equation}
D(0)= \left({\partial Q\over \partial f_{ext}}\right)_T
={1\over \omega_0^2} 
\end{equation}
where 
\begin{equation} 
\omega_0^2=\omega_{\infty}^2-\Pi (0).
\end{equation}
Equivalently, employing Eqs.(31) and (33) 
\begin{equation} 
\omega_{\infty}^2-\omega_0^2=\left({2\over \pi}\right)
\int_0^{\infty} \Re e \ \Gamma(\omega+i0^+)d\omega
\end{equation}
as previously derived in Eq.(19) from purely thermodynamic 
reasoning.

If the dynamical oscillator propagator is presumed to be retarded, 
it too must satisfy a dispersion relation   
\begin{equation}
D(\zeta)=\left({2 \over \pi}\right) 
\int_0^{\infty} {\omega\  \Im m \ D(\omega+i0^+)
\over (\omega^2-\zeta^2)}\ d\omega.
\end{equation}
If the oscillator is stable, then causality is obeyed with the dissipative 
condition  \begin{math}\Im m \ D(\omega+i0^+)\ge 0\end{math}. The 
response function \begin{math} D(\zeta ) \end{math} is then analytic 
in the upper half \begin{math} \Im m\ \zeta > 0  \end{math} of the 
complex frequency plane.
  
On the other hand if the oscillator is unstable, then at least one pole 
at \begin{math} \zeta =i\varpi  \end{math} appears in the upper-half plane 
and the dispersion relation  Eq.(35) for the dynamical oscillator 
propagator is no longer valid. The occurrence of an instability in the 
system can be viewed as inducing the non-analyticity of the dynamical 
oscillator propagator given in Eq.(28). To locate an unstable oscillator 
pole in the upper half plane at \begin{math} \zeta -i\varpi \end{math}, 
we look for a zero in the denominator of Eq.(28); i.e. 
\begin{equation}
\omega_\infty ^2-(i\varpi )^2-\Pi(i\varpi )=0.
\end{equation}
From Eqs.(31) and (36) we seek a solution for 
\begin{math} \varpi >0 \end{math} of the equation  
\begin{equation}
\omega^2_{\infty}+\varpi^2 =\left({2\over \pi}\right)\int_0^{\infty} 
{\omega^2\Re e \ \Gamma(\omega+i0^+)\over (\omega^2+ \varpi^2)}d\omega.  
\end{equation}
As a function of imaginary frequency in the range  
\begin{math} 0< \varpi < \infty \end{math}, it is evident 
that the right hand side of Eq.(36) is smoothly and monotonically  
decreasing from \begin{math} \Pi(0) \end{math} to zero, 
while the left hand side of Eq.(36) is smoothly and  
monotonically increasing from \begin{math} \omega_\infty ^2 \end{math} 
to infinity. If \begin{math} \Pi(0)>\omega_\infty ^2  \end{math}, 
then the two smooth curves intersect at precisely one value 
of \begin{math} \varpi  \end{math}. It is easily proved that 
\begin{math} \zeta=i\varpi  \end{math} is an isolated zero in the 
denominator of Eq.(28). Thus \begin{math} D(\zeta )\end{math} has 
an isolated pole in the upper-half plane. The condition for this 
pole to exist, i.e. 
\begin{equation}
\omega^2_{\infty}<\left({2\over \pi}\right)\int_0^{\infty}
\Re e \ \Gamma(\omega+i0^+)d\omega,
\end{equation}
is the same as the existence of the oscillator instability.

\section {Optical Cavity Instability} 

Consider an electromagnetic mode in an empty optical cavity
\cite{27,28,29,30,31,32}. The 
electric field of the mode 
\begin{math} {\bf E}_\infty ({\bf r}) \end{math}
may be related to the oscillator displacement 
\begin{math} Q \end{math} of the mode by equating the mean energy stored 
in the electric field to the potential energy of the oscillator; i.e. 
\begin{equation}
\left({1\over 8\pi}\right)
\int_{cavity} |{\bf E}_\infty ({\bf r})|^2 d^3{\bf r}
=\left({1\over 2}\right)\omega_\infty ^2Q^2.
\end{equation}
In order to implement Eq.(39), one may choose 
\begin{equation}
 {\bf E}_\infty ({\bf r})= 
\sqrt{4\pi}\ \omega_{\infty} Q\ {\bf e}_{\infty}(\bf r)
\end{equation}
where \begin{math}  {\bf e}_{\infty}(\bf r)  \end{math} 
is normalized according to  
\begin{equation}
\int_{cavity} |{\bf e}_\infty ({\bf r})|^2 d^3{\bf r}
=1.
\end{equation}
Let us further suppose that cavity has been designed so that 
the mode electric field is localized in a spatial region of 
length scale \begin{math} L  \end{math} about the origin. 
Then  
\begin{equation}
{\bf e}_\infty ({\bf 0})=\left({\bf n}\over L^{3/2}\right),
\end{equation}
where \begin{math} {\bf n} \end{math} is a unit vector; 
\begin{math} ({\bf n\cdot n})=1 \end{math}.

Let us finally suppose that a mesoscopic object is placed within 
the mode neighborhood. The mesoscopic object interacts with the 
electric field via the electric dipole moment 
\begin{math} {\bf p} \end{math}.
The interaction Hamiltonian between the mesoscopic object and 
the optical mode oscillator then takes the form 
\begin{equation}
H_{int}=-{\bf p\cdot E}_{\infty}({\bf 0})=-fQ.
\end{equation}
so that the force on the oscillator coordinate is due to the electric 
dipole moment. The force is   
\begin{equation}
f=\sqrt{(4\pi/L^3)}\ \omega_{\infty}({\bf n\cdot p})=
\sqrt{(4\pi/L^3)}\ \omega_{\infty}p.
\end{equation}
Eq.(44) follows from Eqs.(40), (42) and (43). 

If the oscillator were driven by an external force at 
frequency \begin{math} \omega \end{math}, then from 
Eqs.(25) and (26) the damping force and displacement 
at frequency \begin{math} \omega  \end{math} obey 
\begin{equation}
\delta f_{\omega }=\Pi(\omega +i0^+)\delta Q_\omega .
\end{equation}
The electric dipole moment response to the electric field defines 
the polarizability 
\begin{math} \alpha (\zeta ) \end{math}
of the mesoscopic object via 
\begin{equation}
\delta p_\omega = \alpha (\omega +i0^+ )
{\bf n\cdot}{\delta \bf  E}_{\infty \omega }({\bf 0}).
\end{equation}
Equivalently, 
\begin{equation}
\delta p_\omega = \alpha (\omega +i0^+ )
\sqrt{(4\pi /L^3)}\ \omega_\infty \delta Q_\omega .
\end{equation}
where Eqs.(40), (42) and (46) have been employed. 
From Eqs.(44), (45) and (47), one can relate the self energy 
part of the mode propagator to the polarizability of the mesoscopic 
object; The relationship is  
\begin{equation}
\Pi(\zeta)=\left({4 \pi\omega_{\infty}^2\over L^3}\right)\alpha(\zeta).
\end{equation}
The condition for the oscillator stability 
\begin{equation}
\omega_0^2=\omega_\infty^2-\Pi(0) > 0,
\end{equation} 
may be written in terms of the static polarizability  
\begin{equation}
4\pi \alpha (0) < L^3\ \ \ ({\rm stable\ oscillator}).
\end{equation}
For a very highly polarizable mesoscopic object, say a single domain 
ferroelectric grain \cite{21,33,34}, the photon oscillator can exhibit 
an instability. The general nature of the non-linear forces 
which eventually stabilize the oscillator must come into play.

\section{Super-Radiant Phase}

The problem of how an oscillator reaches a new thermodynamic 
equilibrium position in the ``unstable'' regime can be discussed 
only in the context of a specific microscopic environmental Hamiltonian 
\begin{math} H \end{math} which enters into Eq.(7). An often discussed 
model for a photon oscillator employs \begin{math} N \end{math} 
identical ``two level atoms''\cite{35,36,37,38} in the thermal environment. 
The Hamiltonian is
\begin{equation}
H = -\left({\epsilon \over 2}\right)\sum_{j=1}^N\sigma_{zj}
=-\epsilon S_z, 
\end{equation}
where \begin{math}\epsilon \end{math} is the energy difference between 
the single atom ground state and the excited state, 
\begin{math} 
{\bf \sigma }_j=(\sigma_{xj},\sigma_{yj},\sigma_{zj}) 
\end{math} 
are the Pauli matrices for the \begin{math} j^{th}  \end{math} 
two level atom and 
\begin{math} 
{\bf S}=(1/2)\sum_{j=1}^N {\bf \sigma }_j
\end{math} 
is formally the ``total spin'' of the mesoscopic object.
The object is placed in the region of the cavity where the 
electric fields of the photon oscillator are concentrated. 
The electric field is presumed to interact with the atomic 
system via the total electric dipole moment component in the 
\begin{math} {\bf n}  \end{math} direction; i.e.
\begin{equation}
p=\mu \sum_{j=1}^N\sigma_{xj}=2\mu S_x, 
\end{equation}
where \begin{math} \mu \end{math} is the dipole moment matrix 
element. Using Eqs.(44) and (52) the force exerted on the photon 
oscillators by the dipole moment of the atomic system is  
\begin{equation}
f=\sqrt{16 \pi\over L^3} \ (\omega_{\infty}\mu S_x) .
\end{equation}
One is now in a position to compute the free energy 
\begin{math} {\cal G}(Q,T) \end{math} 
for {\em all} domains (stable or unstable) of the (now non-linear) 
oscillator. 

Substituting Eqs (51) and (53) into Eq.(7) and taking the trace 
over the formal spin degrees of freedom yields  
$$
{\cal G}(Q,T)=\omega_\infty^2 (Q^2/2)+k_BT\ \ln 
\ \cosh{\left[\epsilon\over 2 k_BT\right]}-
$$
\begin{equation}
k_BT\ \ln \cosh\left[{1\over 2k_BT}
\sqrt{{\epsilon^2 }+
\left({16\pi \omega_\infty^2 \mu^2 Q^2 \over L^3}\right)}
\ \right]. 
\end{equation}
Using Eqs (11) and (54), the shift in the frequency of the 
photon oscillators from \begin{math}\omega_\infty\end{math} 
down to the temperature dependent  
\begin{math}\omega_0\end{math} is found to be 
\begin{equation}
\omega_0(T) = \omega_\infty \sqrt{1-\left(
8\pi N\mu^2\over  L^3 \epsilon\right)
\tanh\left(\epsilon \over 2k_BT\right)}\ .
\end{equation} 
The stability of the oscillator requires 
\begin{math} \omega_0(T)>0 \end{math}, 
which in virtue of Eq.(55) reads 
\begin{equation}
T>T_c \ \ ({\rm stable\ oscillator}),
\end{equation}
where the critical temperature for stability obeys
\begin{equation}
\tanh\left(\epsilon \over 2k_BT_c\right)= 
\left( L^3 \epsilon\over 8\pi N \mu^2\right).
\end{equation} 

The shift in the equilibrium position required to stabilize 
the oscillator when \begin{math} T<T_c   \end{math} is found from 
the free energy minimum condition 
\begin{math}\left(\partial {\cal G}/ \partial Q\right)_T=0 \end{math}. 
The resulting equilibrium position \begin{math} Q_e \end{math} obeys 
the transcendental equation  
$$
tanh\left( {\epsilon \over 2k_BT}
\sqrt{1+\left({16\pi \omega_\infty^2 \mu^2 Q_e^2 \over L^3 \epsilon ^2}
\right)}\right)
$$
\begin{equation}
={L^3\epsilon\over8 \pi \mu^2{\cal N}} 
\sqrt{1+\left({16\pi \omega_\infty^2 \mu^2 Q_e^2 \over L^3 \epsilon ^2}
\right)}.
\end{equation}

In Fig.\ 1, with \begin{math} y=(\epsilon/2k_BT_c)=0.1 \end{math},  
we show the following: (i) Above the critical temperature  
\begin{math} x=(T/T_c )>1  \end{math} the renormalized frequency 
of the stable oscillator 
\begin{math} \omega_0(T)=\nu (x,y)\omega_\infty  \end{math} is lowered as 
the temperature is lowered. (ii) Below the critical temperature 
\begin{math} x=(T/T_c )<1  \end{math}, the oscillator position shift 
\begin{math} Q_e(T)=(\sqrt{Nk_BT_c}/\omega_\infty )\zeta (x,y) \end{math}
(required to restore stability) increases as the temperature is lowered. 
Thus as the temperature is lowered from above the critical temperature, 
the oscillator frequency decreases until the ``soft mode'' has vanishing 
frequency at the critical temperature. The transition to an ordered phase 
begins as the temperature is lowered below the critical temperature. The 
oscillator seeks a new equilibrium position 
\begin{math} Q_e(T<T_c) \end{math} displaced from the old equilibrium 
position of \begin{math} Q=0 \end{math}. The further the temperature is 
lowered, the more growth is exhibited in \begin{math} Q_e \end{math}.

In Fig.\ 2 the free energy is plotted as a function of the displacement 
of the oscillator for two values of the temperature. For 
\begin{math} T>T_c \end{math}, the potential has a single minimum at 
\begin{math} Q=0 \end{math}. For \begin{math} T<T_c \end{math}, the 
potential has a double minimum at \begin{math} Q=\pm Q_e(T) \end{math}.
The oscillator will choose one of these minima in the super-radiant phase.

\begin{figure}[htbp]
\begin{center}
\mbox{\epsfig{file=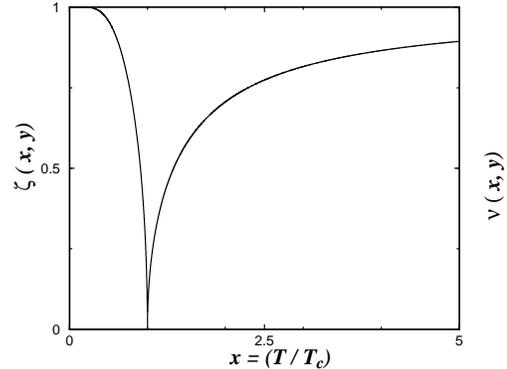,height=50mm}}
\caption{For $y=(\epsilon /k_BT_c)=0.2$, the renormalized 
oscillator frequency  $\omega_0=\omega_\infty \nu $ is 
shown above the critical temperature $x=(T/T_c )>1 $. 
The shifted equilibrium position 
$Q_e=(\sqrt{Nk_BT_c}/\omega_\infty )\zeta $ is plotted below the 
critical temperature $x=(T/T_c )<1$.} 
\label{fig1ajp}
\end{center}
\end{figure} 

\medskip

\begin{figure}[htbp]
\begin{center}
\mbox{\epsfig{file=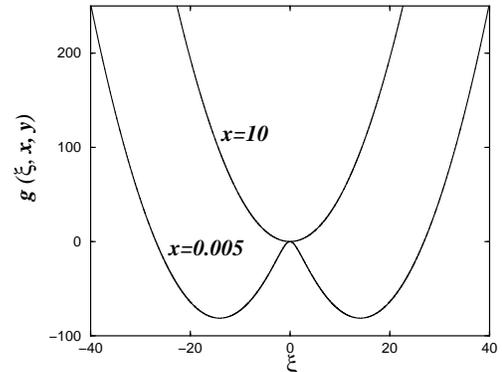,height=50mm}}
\caption{For $y=(\epsilon /k_BT_c)=0.2$, the free energy of the photon 
oscillator  ${\cal G}(Q,T)=Nk_BT g(\xi ,x,y)$ is plotted as a function 
of $Q=(\sqrt{Nk_BT}/\omega_\infty ) \xi $ for two  
values of $x=(T/T_c)$, above the critical temperature  $x>1$ and below 
the critical temperature $x<1$.}
\label{fig1ajp}
\end{center}
\end{figure}

If a highly polarizable mesoscopic grain is 
coupled strongly to the transverse electromagnetic 
field via the electric dipole moment, then there is  
a transition from a normal phase (\begin{math}T>T_c \end{math}) 
to a super radiant phase (\begin{math}T<T_c \end{math}). 

\section {Conclusions}

It has been shown that any simple harmonic oscillator subject 
to a random force (producing damping) can be made unstable if 
the induced damping of the oscillator is sufficiently strong.
If \begin{math} \omega_\infty  \end{math} is the frequency of 
the unperturbed oscillator and if 
\begin{math} \Pi(\zeta ) \end{math} in Eq.(31) 
is the self energy induced by the damping coefficient
\begin{math} \Re e \Gamma(\omega +i0^+ ) \end{math}, then 
the precise condition for the instability is that 
\begin{math} \Pi(0)>\omega_\infty ^2 \end{math}

It is possible to observe this instability for a mesoscopic grain 
which is highly polarizable. The grain must be placed in an optical 
cavity with a mode of spatial extent \begin{math} L \end{math} large 
on the scale of the grain size. The condition for an optical cavity 
mode to go unstable is that 
\begin{equation}
4\pi \alpha (0)>L^3 \ \ \ ({\rm unstable\ cavity\ mode}), 
\end{equation} 
where \begin{math} \alpha(\omega +i0^+ ) \end{math} is the dynamic 
polarizability of the grain. The inequality in Eq.(59) can surely be 
reached with a single domain highly polarizable ferroelectric grain.  

The manner in which the oscillator is eventually stabilized (when
Eq.(59) holds true) is not a trivial problem. The example of a grain 
which consists of ``two-level atoms'' was worked out in detail. The 
normal and super-radiant free energies were exhibited in Fig.\ 2. 
When the frequency of the oscillator mode softens towards zero as 
\begin{math} T\to T_c+0^+  \end{math}, the equilibrium position of the 
oscillator shifts to a nonzero value \begin{math} Q_e(T<T_c) \end{math}. 
For two level atoms, the details of the super-radiant transition 
can be computed.  

It is to be hoped that the theoretical considerations here put forth 
for ferroelectric grains will soon be put to laboratory tests.

\end{document}